\documentclass{article}
\usepackage{spconf,amsmath,graphicx}

\usepackage{amsfonts,amssymb} 
\usepackage{enumitem}
\setlist{nosep, leftmargin=14pt}
\usepackage{CJK}
\usepackage{mwe} 

\title{DECOR-NET: A COVID-19 Lung Infection Segmentation Network Improved by Emphasizing Low-level and Decorrelated Features}
\name{Jiesi Hu$^{\textsuperscript{1} \textsuperscript{2}}$ \qquad Yanwu Yang$^{\textsuperscript{1} \textsuperscript{2}}$ 
\qquad Xutao Guo$^{\textsuperscript{1} \textsuperscript{2}}$ Ting Ma$^{\textsuperscript{1} \textsuperscript{2} \textsuperscript{3} \textsuperscript{4}}$\thanks{Corresponding author: Ting Ma(tma@hit.edu.cn)}}

\address{$^{\textsuperscript{1}}$ Electronic \& Informatin Engineering School, Harbin Institute of Technology (Shenzhen), Shenzhen, China\\
    $^{\textsuperscript{2}}$ Peng Cheng Laboratory, Shenzhen, China \\
    $^{\textsuperscript{3} }$ Guangdong Provincial Key Laboratory of Aerospace Communication and Networking Technology \\
    Harbin Institute of Technology (Shenzhen), Shenzhen, China\\
    $^{\textsuperscript{4}}$ International Research Institute for Artifcial Intelligence\\
    Harbin Institute of Technology (Shenzhen), Shenzhen, China}

\begin{document}
%
\maketitle
\begin{abstract}\
Since 2019, coronavirus Disease 2019 (COVID-19) has been widely spread and posed a serious threat to public health.
Chest Computed Tomography (CT) holds great potential for screening and diagnosis of this disease.
The segmentation of COVID-19 CT imaging can achieves quantitative evaluation of infections and tracks disease progression.
COVID-19 infections are characterized by high heterogeneity and unclear boundaries, so capturing low-level features such as texture and intensity is critical for segmentation.
However, segmentation networks that emphasize low-level features are still lacking.
In this work, we propose a DECOR-Net capable of capturing more decorrelated low-level features.
The channel re-weighting strategy is applied to obtain plenty of low-level features and the dependencies between channels are reduced by proposed decorrelation loss.
Experiments show that DECOR-Net outperforms other cutting-edge methods and surpasses the baseline by 5.1\% and 4.9\% in terms of Dice coefficient and intersection over union.
Moreover, the proposed decorrelation loss can improve the performance constantly under different settings.
The Code is available at https://github.com/jiesihu/DECOR-Net.git.
\end{abstract}
\section{Introduction}
\label{sec:intro}

\begin{figure}[t]
\centering 
\includegraphics[width=6cm]{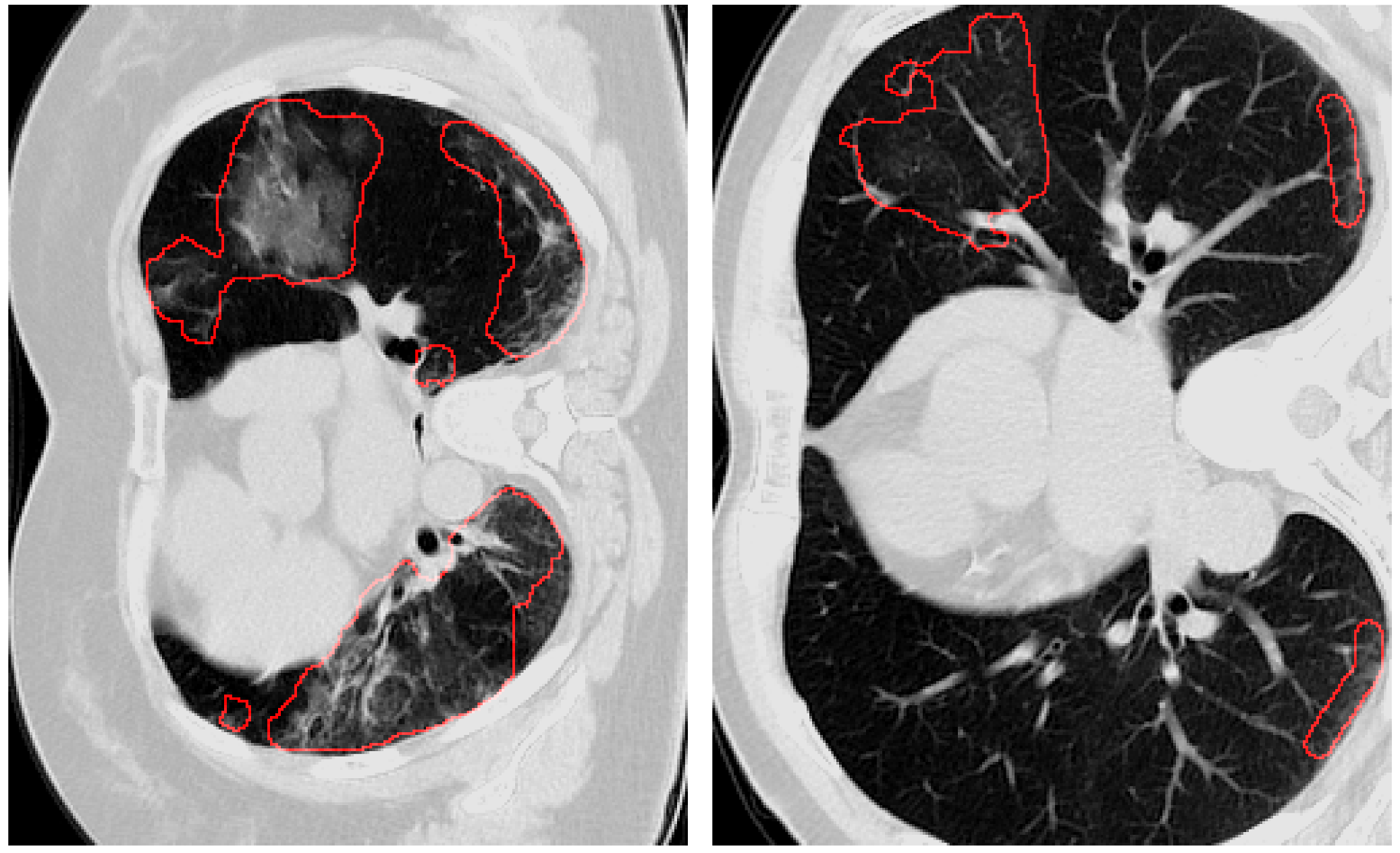} 
\caption{Examples of COVID-19 infected regions from COVID-19 Challenge dataset \cite{roth2021rapid}. The infection area is delineated by the red line.}
\label{Figinf} 
\end{figure}

The coronavirus disease (COVID-19) has spread rapidly to countries around the world since 2019.
In March 2020, it was declared by World Health Organization (WHO) as a pandemic \cite{roosa2020real}.
The pandemic has presented severe challenges to routine daily life, the global economy, and general public health.
However, medical resources for COVID-19 are very limited compared to demand.
As a complement to the RT-PCR test, computed tomography (CT) imaging of the human lung is considered a tool for diagnosing and monitoring COVID-19 infection.
Several patterns such as ground-glass opacity, multifocal patchy consolidation, and crazy-paving pattern have been declared as a result of COVID-19 infection \cite{liu20202019}.
However, the segmentation of COVID-19 infection is still challenging due to the characteristics of high variation in pattern, shape, location, and blurred boundary as shown in figure \ref{Figinf}.
  
In the field of medical image segmentation, deep learning networks, such as U-net \cite{ronneberger2015u}, Res-UNet \cite{diakogiannis2020resunet}, and transformer-based models \cite{cao2021swin},
have been proposed and achieved promising results.
Recently, many networks were proposed for COVID-19 infection segmentation. 
Inf-Net \cite{fan2020inf} took edge information on the lung as one of the supervision and introduced specific attention and multi-scale mechanisms.
\cite{karthik2022contour} utilized the information from boundary and shape to precisely capture infected tissues.
These studies tried to take advantage of edge and shape information, but this may only provide limited help due to the blurred boundary of the infections. 
To alleviate the problem of insufficient data, semi-supervised \cite{liu2022ccat}, self-supervised \cite{fung2021self}, and weakly supervised \cite{laradji2021weakly} were applied in the segmentation of COVID-19 infection.
Radiologists mainly rely on low-level features such as texture, line, and intensity to identify infections, due to the heterogeneity of the shape and location \cite{ng2020imaging}.
However, there is still a lack of models that emphasize low-level features, which is crucial for identifying the pattern of the infection.

To detect subtle differences in low-level features, we propose DECOR-NET, a network that adds more decorrelated features in the shallow layers.
We apply the channel-reweighting strategy to increase the number of channels in the early layer of the network and add the proposed decorrelation loss to ensure the diversity of low-level features.
Compared with  other decorrelation method \cite{cogswell2015reducing, wang2020orthogonal}, the proposed decorrelation loss does not have the undesired effect of weight decay 
and directly decorrelate the feature map instead of the weight.
Note that our method does not add any extra parameters to the network.

Our contributions in this work are threefold:
(1) We introduce a strategy to improve the COVID-19 infection segmentation by utilizing plenty of decorrelated low-level features.
(2) A novel loss function is proposed to reduce the correlation between feature channels.
(3) Extensive experiments showed that with a relatively small size our proposed network outperforms most existing networks.

\begin{figure}[htb]
\centering 
\includegraphics[width=8.5cm]{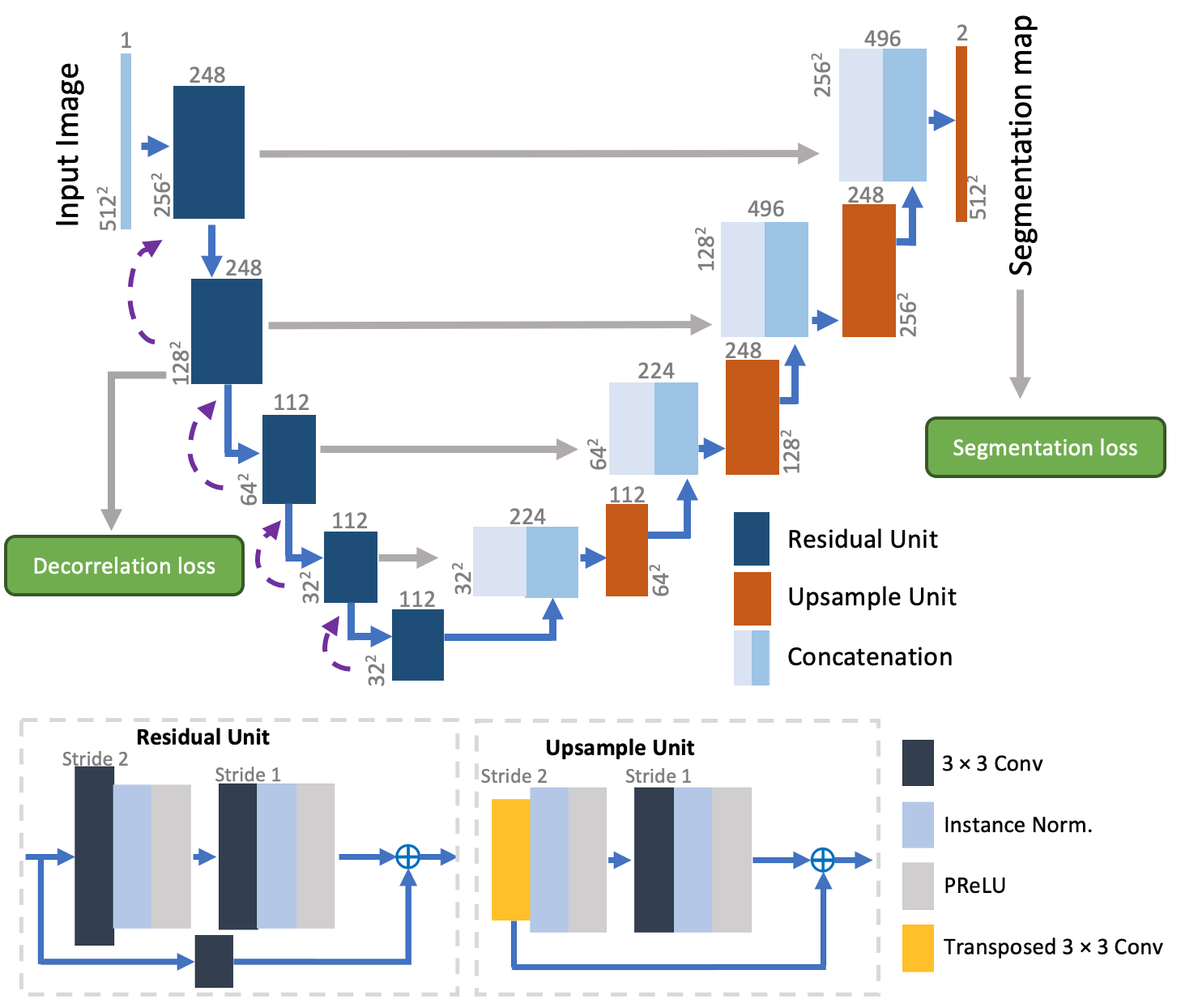} 
\caption{Overview of the Network. The gray number around the block is the output size of each unit. The purple dashed arrows illustrate the channel re-weighting strategy.}
\label{Fig.main1} 
\end{figure}

\section{Method}
\label{sec:method}
The detailed architecture of the proposed network is presented in Figure \ref{Fig.main1} and Res-UNet\cite{diakogiannis2020resunet}, a widely used network for medical image segmentation, is selected to be the baseline of our model. 
We mainly made two changes on Res-UNet, which are applying the channel re-weighting strategy and adding the proposed decorrelation loss on the output feature maps of all encoder units.

\subsection{Channel re-weighting}
\label{ssec:channel}
\cite{zeiler2014visualizing} shows that the shallow layers of the network capture low-level features such as texture, edge, and intensity, while the deep layers preserve the semantic information.
Usually, Res-UNet has 5 layers and the channel number of these five layers is 32, 64, 128, 256, and 512 in order.
However, this regular setting is not suitable for the COVID-19 infection segmentation task which desires relatively more low-level features.
Thus, the channel re-weighting strategy strengthens the model's ability to perceive low-level features by directly increasing the number of channels in the shallow layers of the network.
In addition, to keep the size of the parameter unchanged, it reduces the channel's number of deep layers simultaneously. 
In other words, it transfers parameters from deep layers to shallow layers.
After trying different settings, we changed the channel setting of Res-UNet from (32, 64, 128, 256, 512) to (248, 248, 112, 112, 112).
We retain the number of encoder units in the network to preserve a large field of view and some high-level semantic information to identify lobe regions.

\begin{figure}[htb]
\centering 
\includegraphics[width=8.5cm]{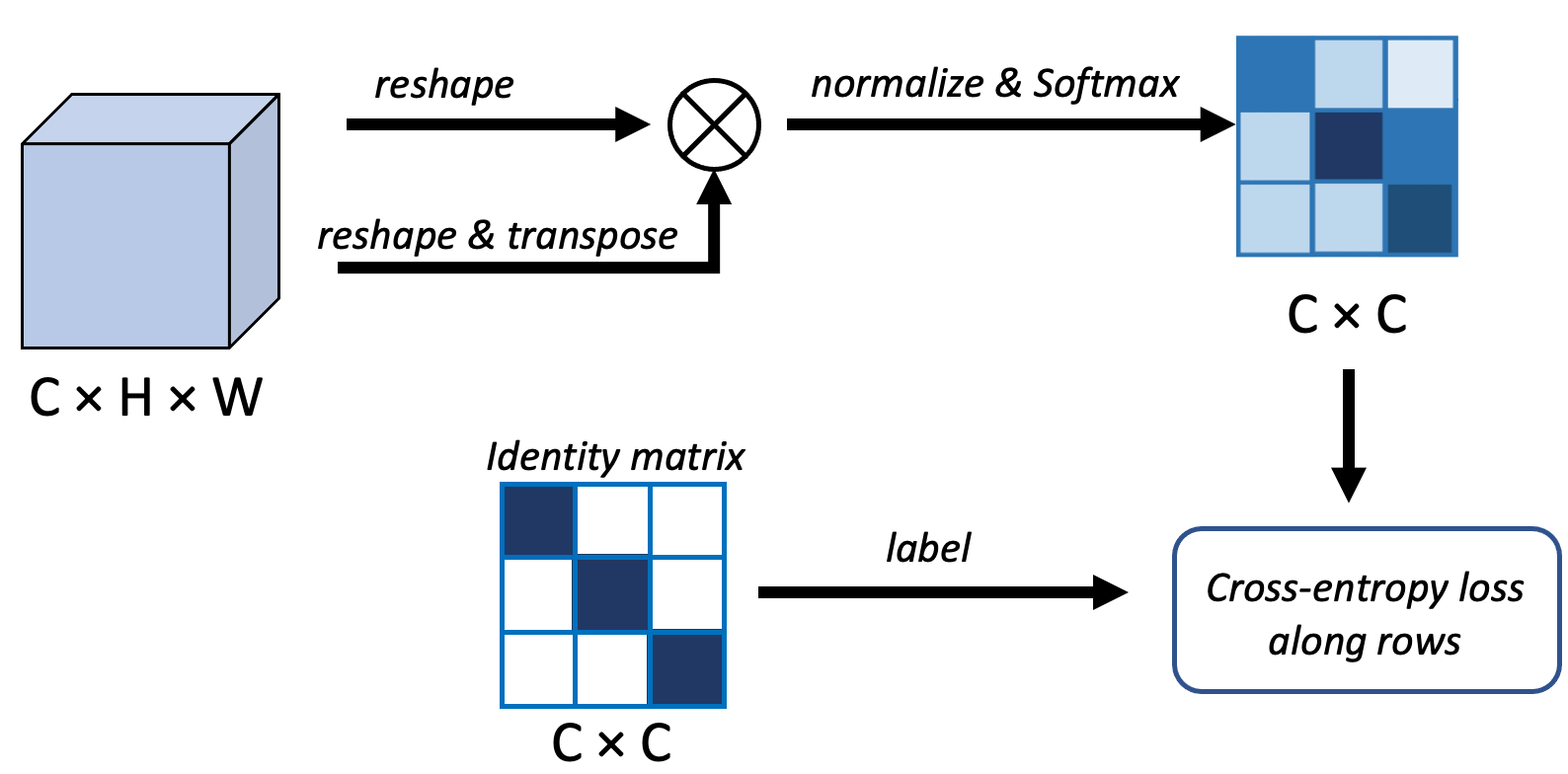} 
\caption{The pipeline of computing decorrelation loss.}
\label{Fig.main2} 
\end{figure}

\subsection{Decorrelation loss}
\label{ssec:decorloss}
Each channel on the feature map contains a specific representation of the input.
Channel re-weight strategy has increased many channels in shallow layers, which may result in redundancy of learned low-level features.
To ensure the diversity of the low-level features, the decorrelation loss (Decor loss) is proposed.

The pipeline of computing the decorrelation loss is shown in Figure \ref{Fig.main2}.
Refer to the channel interdependency of \cite{fu2019dual}, we calculate the channel correlation map $C\in \mathbb{R} ^{C\times C}$ from the original features $H\in \mathbb{R} ^{C\times H\times W}$:

\begin{equation}
  c_{i,j} = \sum_{h}^H\sum_{w}^W h_{i}^{h,w}h_{j}^{h,w} 
\end{equation}
where $c_{i,j}$ denotes the the value of the $i$th row and the $j$th column on the correlation map $C$.
$h_{i}^{h,w}$ denotes the the value of the $h$th row and the $w$th column on the $i$th channel of the feature map $H$.
For simplicity, we can also reshape $H\in \mathbb{R} ^{C\times H\times W}$ to $H\in \mathbb{R} ^{C\times HW}$:
\begin{equation}
  c_{i,j} = H_{i}\cdot H_{j}
\end{equation}
Then, we apply a softmax function to obtain the probability map $X\in \mathbb{R} ^{C\times C}$:
\begin{equation}
  x_{i,j} = \frac{exp(\frac{c_{i,j}}{z_{i}})}{\sum_{k}^C exp(\frac{c_{i,k}}{z_{i}})}
\end{equation}
where $z_{i}$ denotes the normalization term and is defined as the largest value in the $i$th row of $C$. 
Adding $z_{i}$ can prevent the model from directly enlarging the scale of $C$ to reduce the loss. 
Note that we treat the selected maximum value as constant when computing the gradient.
Finally, we add cross-entropy loss, hoping that the channel only relates to itself:
\begin{equation}
  L_{decor} = -\sum_{i}^C log(x_{i,i})
\end{equation}
To understand this further, consider the gradient of the loss with respect to a particular activation $h_{a}^{h',w'}$.
\begin{equation}
\begin{split}
  \frac{\partial L_{decor}}{\partial h_{a}^{h',w'}} = \sum_{i\neq a}(\frac{x_{a,i}}{z_{a}}+
  \frac{x_{i,a}}{z_{i}})h_{i}^{h',w'}\\
  +\frac{2(x_{a,a}-1)}{z_{a}}h_{a}^{h',w'}  
\end{split}
\end{equation}
There are two terms in the above formula. 
The first term illustrates the influence of other channels on channel $a$ which allows channel $a$ to decorrelate with others. 
The second term makes channel $a$ enhance its value.
Compared with \cite{cogswell2015reducing}, our proposed decorrelation loss will not have the undesired effect of weight decay,
because increasing or decreasing the scale of the weight does not change $X$.

Denoting $\lambda > 0$ as the weight of the decorrelation loss, the overall loss function $L$ is defined as follows:
\begin{equation}
  L = \frac{1}{2}L_{CE}+\frac{1}{2}L_{DC}+\lambda L_{decor}
  \label{eq.L}
\end{equation}
where $L_{CE}$ denotes the binary cross entropy loss and $L_{DC}$ denotes the Dice loss.

\section{Experiments}
\label{sec:experiments}

\begin{table}[!htb]
\centering
\small
\setlength\tabcolsep{2pt}
\caption{Comparison with other state-of-art methods on the test sets.}
\label{Tab1}
\begin{tabular}{c|c|cccc}
\hline

Method& Param.& Dice & IoU& Precision& Recall\\
\hline
U-Net \cite{ronneberger2015u}& 2.637 M&	0.6097&	0.4654&	0.6647&	0.6457\\
Attention U-net \cite{oktay2018attention}&	8.725 M&	0.5883&	0.4500&	0.6449&	0.6195\\
U-net++ \cite{zhou2018unet}&	9.045 M&	0.5977&	0.4580&	0.6702&	0.6169\\
Inf-Net \cite{fan2020inf}&	33.122 M&	0.6129&	0.4689&	0.6504&	0.6508\\
U-net++ (large) \cite{zhou2018unet}&	36.165 M&	0.6053&	0.4664& 0.6672&	0.6372\\
Swin-Unet \cite{cao2021swin}&	41.342 M&	0.5998 & 0.4567 & 0.6423&	0.6502\\
\textbf{DECOR-Net(Ours)}&	6.457 M&	\textbf{0.6378} &	\textbf{0.4940}&	\textbf{0.6788}&	\textbf{0.6799}\\
\hline
\end{tabular}
\end{table}

\subsection{Datasets and Evaluation Metrics}
\label{ssec:metrcs}
We run our experiments on the public COVID-19 Challenge \cite{roth2021rapid} dataset
which contains 199 lung CT volumes of COVID-19 patients including 9704 CT slices of $512\times 512$ size.
After splitting, there are 127 volumes in the training set, 32 volumes in the validation set, and 40 volumes in the test set.

In this study, we adopt the Dice similarity coefficient (DSC), intersection over union (IoU), precision, and recall as evaluation metrics.

\subsection{Implementation details}
\label{ssec:imple}
Our model was implemented in PyTorch using the MONAI framework \cite{cardoso2022monai}.
Stochastic gradient descent (SGD) and Adam optimizer were implemented to optimize the model.
The learning rate was initialized to $1\times10^{-4}$.
Whenever training loss did not decrease by at least $5\times10^{-3}$ within the 30 epochs, the learning rate was reduced by a factor of 5.
All models were trained for 300 epochs.
The validation set was used to select epoch with the best model, while the performances of models were finally evaluated on the test set.
The augmentation methods we applied include random rotation, scaling, elastic deformations, gamma correction, mirroring, and intensity shifting.
The hyper-parameter $\lambda$ for decorrelation loss was set to be 0.01.
\begin{table}[!htb]
\centering
\small
\setlength\tabcolsep{2pt}
\caption{Model performances using different decorrelation methods. CR denotes the channel re-weighting strategy.}
\label{Tab.main3}
\begin{tabular}{c|cccc}
\hline
Decorrelation Method& Dice & IoU& Precision& Recall\\
\hline
Only CR&	0.6264&	0.4853&	0.6749&	0.6690\\
CR + Deconv loss\cite{cogswell2015reducing}&	0.6301&	0.4868&	0.6669&	0.6799\\
CR + Ortho loss\cite{wang2020orthogonal}&	0.6298&	0.4862&	0.6702&	\textbf{0.6837}\\
CR + Decor loss (ours)&	\textbf{0.6378}&	\textbf{0.4940}&	\textbf{0.6788}&	0.6799\\
\hline
\end{tabular}
\end{table}
  
\subsection{Comparison with other models}
\label{ssec:compare}
Tabel \ref{Tab1} shows the quantitative results compared with other state-of-art methods.
Our model achieves competitive results on all metrics.
We experimented without loading the pre-trained model.
All models are under the same training framework, except that Inf-net uses its published framework.

\begin{figure}[htb]
\begin{minipage}[b]{.48\linewidth}
  \centering
  \centerline{\includegraphics[width=4.3cm]{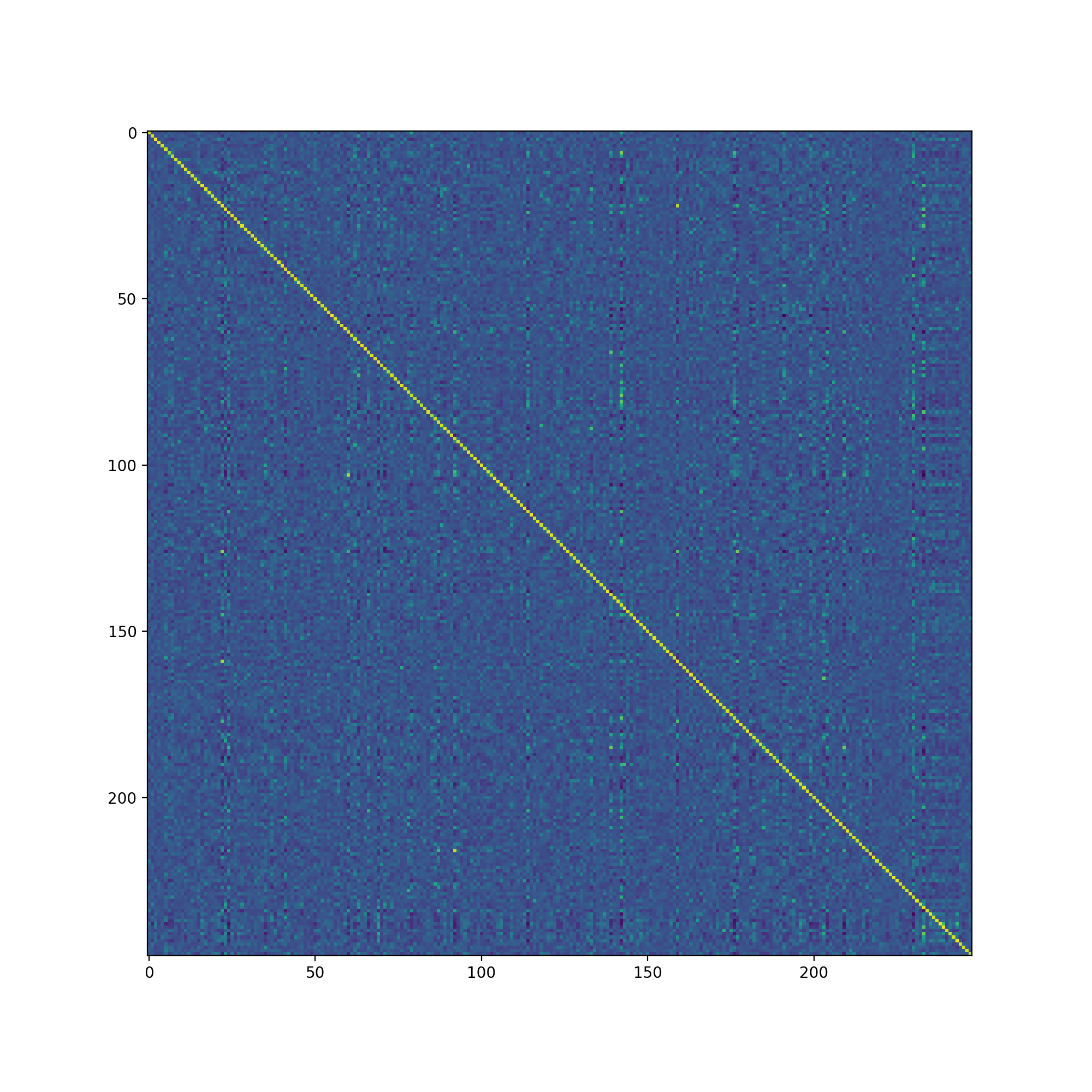}}
  \centerline{(a)}\medskip
\end{minipage}
\hfill
\begin{minipage}[b]{0.48\linewidth}
  \centering
  \centerline{\includegraphics[width=4.3cm]{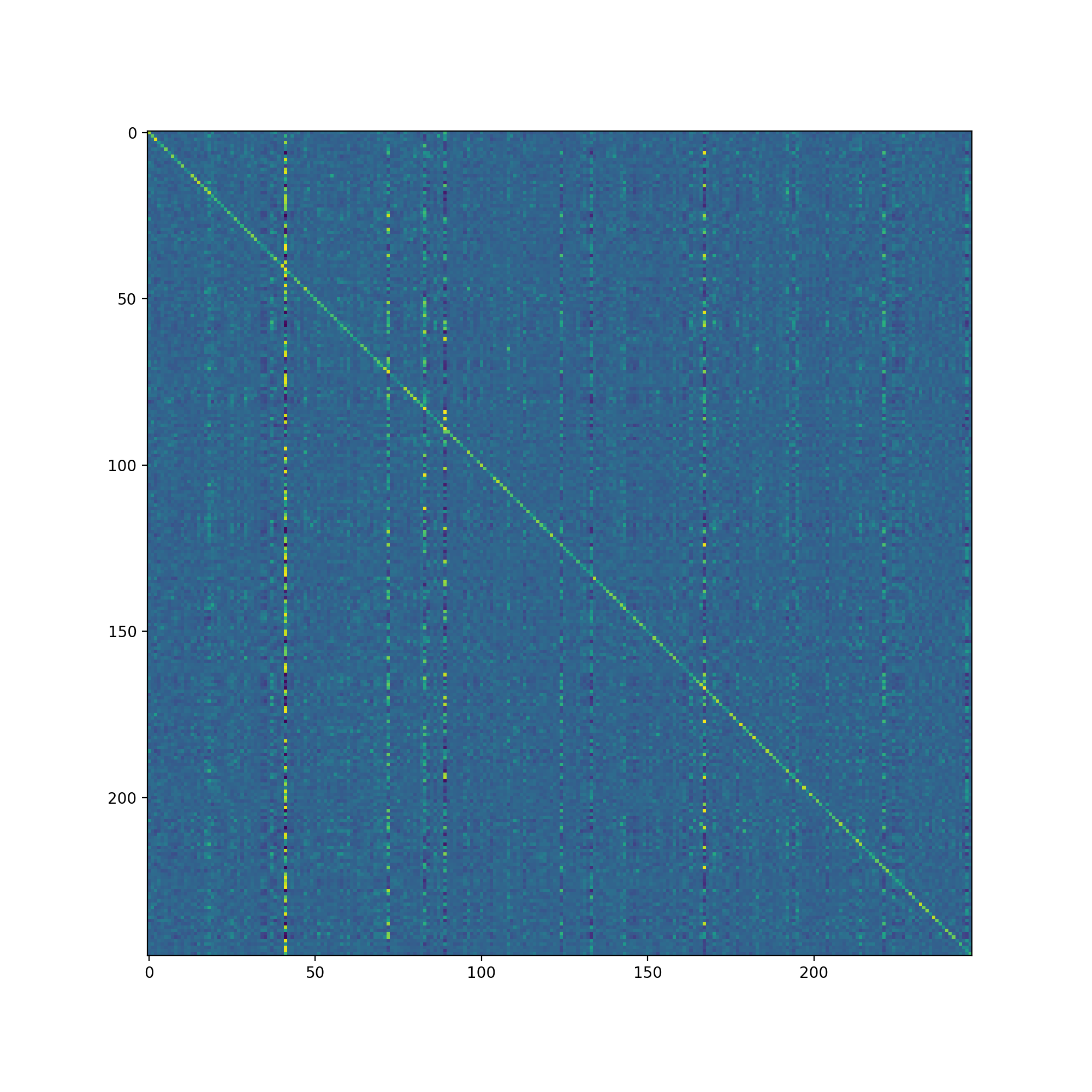}}
  \centerline{(b)}\medskip
\end{minipage}
%
\caption{(a) and (b) show the probability matrix $X$ with and without Decor loss for the output of the second layer.}
\label{fig:3}
\end{figure}
\begin{figure}[htb]
\centering 
\includegraphics[width=8.5cm]{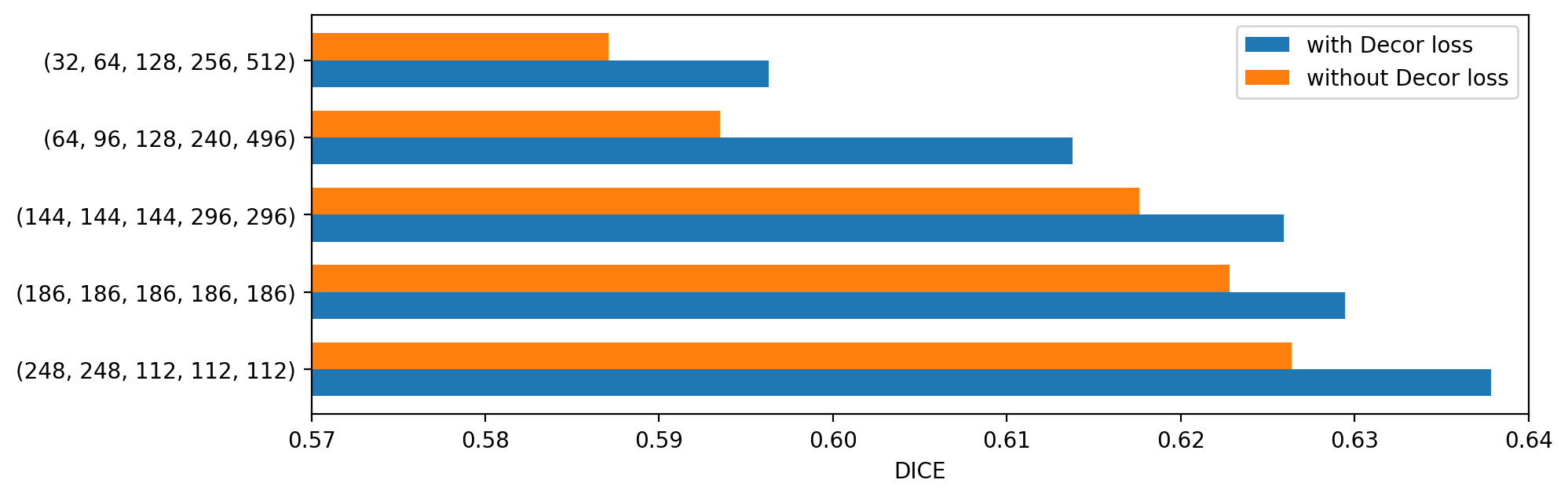} 
\caption{Performances under different channel settings.}
\label{fig:4} 
\end{figure}

In table \ref{Tab.main3}, we compare the proposed decorrelation loss with two other decorrelation methods.
Hyper-parameter tuning is performed for each method.
The proposed decorrelation loss still achieves the best performance.
Figure \ref{fig:3} shows the average probability map $X$ over all slices in the validation set, 
indicating that decorrelation loss can make the values on the diagonal larger and reduce the interdependences between channels.
Figure \ref{fig:4} shows that adding decorrelation loss can stably improve the model performance under different channel settings.
In addition, the model performance becomes better as the number of channels in the early layer becomes larger.

\subsection{Ablation study}
The ablation study was performed to demonstrate the effectiveness of our methods.
Results in table \ref{Tab3} show the benefits of the channel re-weighting strategy and the decorrelation loss.
\label{ssec:ablation}
\begin{table}[!htb]
\centering
\small
\setlength\tabcolsep{2pt}
\caption{Ablation study of channel re-weighting strategy (CR) and decorrelation loss.}
\label{Tab3}
\begin{tabular}{cc|c|cccc}
\hline
CR& Decor loss& Param.& Dice & IoU& Precision& Recall\\
\hline
 & & 6.495 M&	0.5871&	0.4450&	0.6371&	0.6488\\
&\checkmark &	6.495 M&	0.5963&	0.4554&	0.6541&	0.6411\\
\checkmark & &	6.457 M&	0.6264&	0.4853&	0.6749&	0.6690\\
\checkmark& \checkmark &	6.457 M&	\textbf{0.6378} &	\textbf{0.4940}&	\textbf{0.6788}&	\textbf{0.6799}\\
\hline
\end{tabular}
\end{table}

\section{Conclusion}
\label{sec:conclusion}
In this paper, we build a DECOR-Net to accurately segment COVID-19 infection from CT volumes.
To capture more discriminating low-level features, we introduce channel re-weighting strategy and a novel decorrelation loss.
The channel re-weighting strategy enlarges the number of channels in the shallow layers 
and the decorrelation loss helps avoid redundancy between channels.
A comprehensive experiment illustrates the proposed network outperforms most cutting-edge methods.
The proposed decorrelation loss can consistently improve model performance under different settings and outperforms other deocorrelation methods.
Moreover, our methods do not increase the model size, which prevents further overfitting and hunger for data.
The proposed network may be widely applicable to lesion segmentation that relies on texture or edge information.

\section{Acknowledgments}
\label{sec:acknowledgments}
This study is supported by grants from the National Natural Science Foundation of China (62276081), the Innovation Team and Talents Cultivation Program of National Administration of Traditional Chinese Medicine (NO:ZYYCXTD-C-202004), Basic Research Foundation of Shenzhen Science and Technology Stable Support Program (GXWD20201230
155427003- 20200822115709001), and The Major Key Project of PCL. 

\bibliographystyle{IEEEbib}
\bibliography{strings,refs}

\end{document}